\documentclass[preprint]{elsarticle}
\usepackage {natbib}
\usepackage {bm}
\usepackage {amsmath,amssymb}
\usepackage {setspace}
\title{The self gravity effect on the orbital stability of Twotinos}
\author[yt]{YUSUKE TSUKAMOTO}
\ead{tukamoto@cfca.jp}
\address[yt]{Department of Astronomy, University of Tokyo,7-3-1 Hongo, 
Bunkyo-ku, Tokyo 113-0033, Japan; tukamoto@cfca.jp}

\begin{document}
\begin{abstract}
We investigate how the self gravity of Twotinos changes its orbital stability using N-body simulations in which
the gravitational interaction between planetesimals are fully calculated.
We show the timescale in which the half of Twotinos becomes unstable, $t_{half}$, obey the formula,
\begin{equation}
t_{half}=4\times10^6(\frac{M_{tot}}{0.1M_\oplus})^{-1}(\frac{m_p}{7.6\times10^{23} g})^{-1}
(\frac{\langle i^2\rangle^{1/2}}{0.002}) (years),\notag
\end{equation}
, if we assume the primordial planetesimals disk have the power low surface mass density, $\Sigma=\Sigma_0\times r^{-2/3}$.
Where $M_{tot}$, $m_p$, $\langle i^2\rangle^{1/2}$ are the total mass of large bodies of Twotinos,
the maximum mass of planetesimals, and the inclination dispersion.
With this formulae, we conclude the total mass of Twotinos is reduced to the order of $0.01~M_{\oplus}$ by
the self gravity and secular perturbation of Planets even if there was huge mass such as several order of the 
earth mass in 1:2 MMR of Neptune at the early age of the solar system. 
These results will invoke reexamination to many previous works explaining the dynamical evolution 
of TNOs.
\end{abstract}

\begin{keyword}{TNOs \sep KBOs \sep planetary system \sep Twotinos \sep mean motion resonance \sep the Solar System
 \sep {\it n}-body \sep GRAPE-6}
\end{keyword}

\maketitle
\section{Introduction}
It is important feature that the highly eccentric planetesimals in the mean motion
resonance (MMR) of Neptune such as Plutinos or Twotinos exist and stay in for the 
age of the solar system. These populations have provided the informations
to establish the dynamical evolution model of the primordial Solar system. 
Malhotra (1995) proposed Resonant sweeping model
in which Neptune migrate several AU capturing primordial TNOs into its 
MMR. This model firstly explain the Pluto's peculiar orbit and successfully predict the
existence of Plutinos, Twotinos, and scattered TNOs. 
But many unexplainable orbital characteristics of TNOs remain. After Malhotra's work, lots of works 
have been done to explain the orbital structures of TNOs more precisely. 
These works are categorized into two types.

The works categorized into first type follow the framework of Malhotra's work and carefully investigate 
how the resulting structure of TNOs is changed by the initial position, migration model 
of planets or the initial distribution of planetesimals disk.
For example, by changing the migration speed of Neptune, Ida {\it et al.} (2000) try to 
explain the ratio of Plutinos to Twotinos. Zhou {\it et al.} (2002) also address
this issue considering the stochastic migration model of Neptune. 
Levison \& Morbidelli (2003) propose a bolder model in which the primordial 
planetesimal disk was assumed to be dynamically cold and truncated around 
35AU. If the mass of Twotinos was about $3~M_{\oplus}$ when the 1:2 MMR of Neptune reached 
35AU and the migration of Neptune after that was properly stochastic,
the outcome can accounts for almost all aspects of TNOs region. 
This model is very attractive because of its strong explanation for the structure of TNOs. 
But the strong stochastic motion of Neptune by planetesimals 
is unlikely because the mass of planetesimals are too small to induce
strong stochastic migration (Murray-Clay \& Chiang 2006).

Nice model (Tsiganis {\it et al.} 2005, Gomes {\it et al.} 2005, Morbidelli {\it et al.} 2005) 
seems to be the most prosperous model to explain the dynamical structure of TNOs. In this model,
the planetesimal disk is also assumed to be truncated around 35 AU and four giant planets 
was formed in more compact configuration than current configuration. During the early migration
phase, Jupiter and Saturn was locked into the mutual 1:2 MMR and this resonance  induced strong
 instability to Uranus and Neptune. By this instability, They were scattered outwards shaking up the outer 
planetesimal disk and reduce their eccentricity by the dynamical frictions from planetesimals. 
This model accounts for
not only all important aspects of TNOs region but also the Late Heavy Bombardment, the Trojan populations
of Jupiter, the eccentricity and inclination of the planets. Nice model is very powerful 
but it assumes a lot of the conditions for the primordial solar system and whether such a dynamical migration
really occur or not is still uncertain.
Recently, Minton \& Malhotra (2009) suggested that the information of the migration process of planets
was graved into the asteroid belt. Thus, what kind of migration occurred in early solar system  
will revealed by the detailed investigations of the asteroid belt.

The works belong to the second type consider the orbital instability after the migration.
The instability of the orbit may change the structure of TNOs.
Many authors (Levison and Stern 1995, Malhotra 1996, Morbidelli 1997, Nesvorn\'y 
and Roig 2000, 2001, Tiscareno \& Malhotra 2009) have investigate the stability of the orbits 
in the resonant region under the perturbation from the Planets. 
Levison \& Stern (1995) studied the stability of 
the orbits in the 2:3 MMR as a function of $A_\sigma$, $A_\omega$ where
 $A_\sigma$, $A_\omega$ are the initial amplitudes of the resonant angle and 
the argument of perihelion.
They found that  $A_\sigma > 120^\circ$ were unstable over $4 \times 10^9$ years 
even if $A_\omega =0 $ and as $A_\omega$ increases, the boundary value of $A_\sigma$ which
separates stable orbits and unstable orbits decreases. As they have shown, the amplitude 
of resonant angle determine the stability of orbits. 

Nesvorn\'y \& Roig (2001) studied the orbital stability of 1:2 MMR using Lyapunov 
characteristic exponent. Their result showed that the orbits which have 
$A_\sigma >30^\circ$, $e=0.3$ $i=5^\circ$ are unstable for the age of 
the Solar System. On the other hand, they pointed out to determine the stable region of intermediate 
eccentricity such as $0.1<e<0.4$ is very difficult because of
its complex chaotic nature. But figure 5 in their paper seems to show 
$A_\sigma >30^\circ$ can be a crude threshold between stable and unstable orbit
for $0.1<e<0.4$ and we use this value as the threshold in this paper.
They also pointed out the additional kick to the amplitude of the resonant angle may play
an important role on the orbital stability of the Twotinos (see, fig 10 of the Nesvorn\'y \& Roig 2001).

Although the gravitational effect of planets on the orbital evolution of TNOs are well investigated,
the study about the mutual gravity effect between planetesimals is very rare.
The gravitational effect on Plutinos by the back ground planetesimals is partially studied in 
Levison and Stern (1995) by the particle-in-box framework. In their work, the gravitational 
effect to Pluto from background particles
which have circular orbits and the size between 1 km - 330 km is considered. 
They conclude that the orbit of Pluto can be stabilized by the effect. 
However, Their investigation is limited and it is important to investigate the mutual gravitational 
effect of resonant TNOs more precisely. 
 
In this paper, we address the issue how the self gravity in the resonant TNOs affects its stability
of the planetesimals with N-body simulation in which the self gravity is fully considered. 
Especially, we investigate Twotinos
because the stable region in the phase space of Twotinos is smaller than that of Plutinos
and Twotinos seem to be destabilized easier than Plutinos and they may affect the structure of TNOs more strongly.   
The analysis about 2:3 MMR will be done in future works.

The numerical method and the initial conditions are described in Section \ref{method}. In
Section \ref{results}, we show our results and estimate the effects of the self gravity 
on the actual Twotinos by extrapolating the results.
We summarize the results in Section \ref{discussion}.

\section{ Method of Calculation and Initial Conditions}
\label{method}
\subsection{Integration Method}
We use the fourth-order Hermite scheme (Makino \& Aarseth 1992) with
hierarchical time-steps (Makino 1991) improved for planetary systems
(Kokubo {\it et al.} 1998 ) for numerical integration of the planetesimals
and Neptune. The gravitational interaction between planetesimals are fully 
considered and the gravitational effect of planetesimals to Neptune or the Sun are neglected.
We use position of the Sun as a origin of the coordinate for
planetesimals and Neptune. 
We use the equation of motion for planetesimals as follow
\begin{eqnarray}
{\bf a}_{i} &=& - \sum_{i \neq j} G m_{j} \frac{ {\bf r_{i j}} }{r_{i j}^{3}} 
- G(m_{i} +M_{\odot})\frac { {\bf r_{i}} } {r_{i}^{3}} \notag\\
&-&G M_{N} 
( \frac { {\bf r_{i N}} }{ r_{i N}^{3} } + \frac { {\bf r_{N}}
}{r_{N}^{3}} ) ~,
\end{eqnarray}
where $M_{\odot},M_{N}, r_{i}$, and $r_{i N}$ are the mass of the Sun,
the mass of  Neptune, position of planetesimal relative
to the Sun and that relative to  Neptune. $r_N$ is position of Neptune relative 
to the Sun.
The gravitational
forces of planetesimals to Neptune is neglected {\it i.e.}, the equation
of motion for Neptune is that of simple two body problem.

We use GRAPE-6 (Makino {\it et al.} 2003)  and GRAPE-6A (Fukushige {\it et
al.} 2005) to calculate the gravitational interaction between
planetesimals. For both the planetesimals and
Neptune, we use the standard time-step criterion (Aarseth 1985) 
\begin{eqnarray}
\Delta t =\sqrt{ \eta \frac{|a||a^{(2)}|+ |\dot{a}|^{2}}
       {|\dot{a}||a^{(3)}|+ |a^{(2)}|^{2}}} ~.
\end{eqnarray}
Where $a,~\dot{a},~a^{(n)}$ are the acceleration and its first order, n-th order time differential.
 
The $\eta$ is selected as the energy error of planetesimals is 
less than 0.003 \% of initial energy of planetesimals. 
Each orbit is integrated with several hundred time-steps on average.
This is sufficient to avoid the artificial effect on the orbit
such as precession.
 
\subsection{Initial Conditions}
We make the initial resonant population by simple resonant capture process by Neptune.
At first, the planetesimals are distributed as a cold disk
whose distributions of eccentricity and inclination 
are both given by Rayleigh distribution with dispersion  
$\langle e^2 \rangle ^{1/2}=2 \langle i^{2} \rangle^{1/2}=0.004$ and its mass is equal to zero.
The number density of the disk is power law of the distance from the Sun and
its exponent is equal to $-3/2$. The inner and outer cutoff of the disk are 
25 AU and 50 AU, respectively. We use 10000 particles to make sufficient
number of resonant particles. 
After migration of Neptune, we picked up the resonant particles for initial conditions.
In our model, only Neptune and Sun are considered and the gravitational effects of other Planets are neglected to investigate
the mutual gravitational effect of TNOs genuinely.

The migration model is similar to that of  Malhotra (1995) and the semi-major axis of Neptune
evolve such as 
\begin{eqnarray}
a(t)=a_f-(a_f-a_i)e^{-t/\tau}
\end{eqnarray} 
where $\tau$ is migration timescale and its value is $5.0 \times 10^6$ years. $a_f,~a_i$ are
the final position and initial position of Neptune and we fix these parameters. the value of 
the parameters are 23 AU and 30 AU, respectively.
As Gomes et al (2004) investigated, the migration mode of Neptune drastically change due to 
the mass and the surface mass density profile of the planetesimal disk and there are many possibilities
for the migration model.
However, we adopt the typical migration model to make our initial conditions for simplicity.
The system is integrated for $1.6\times 10^7$ and Neptune migrate from 23 AU to 29.7 AU
trapping the planetesimals into its resonance.

\begin{figure*}
\includegraphics[width=150mm]{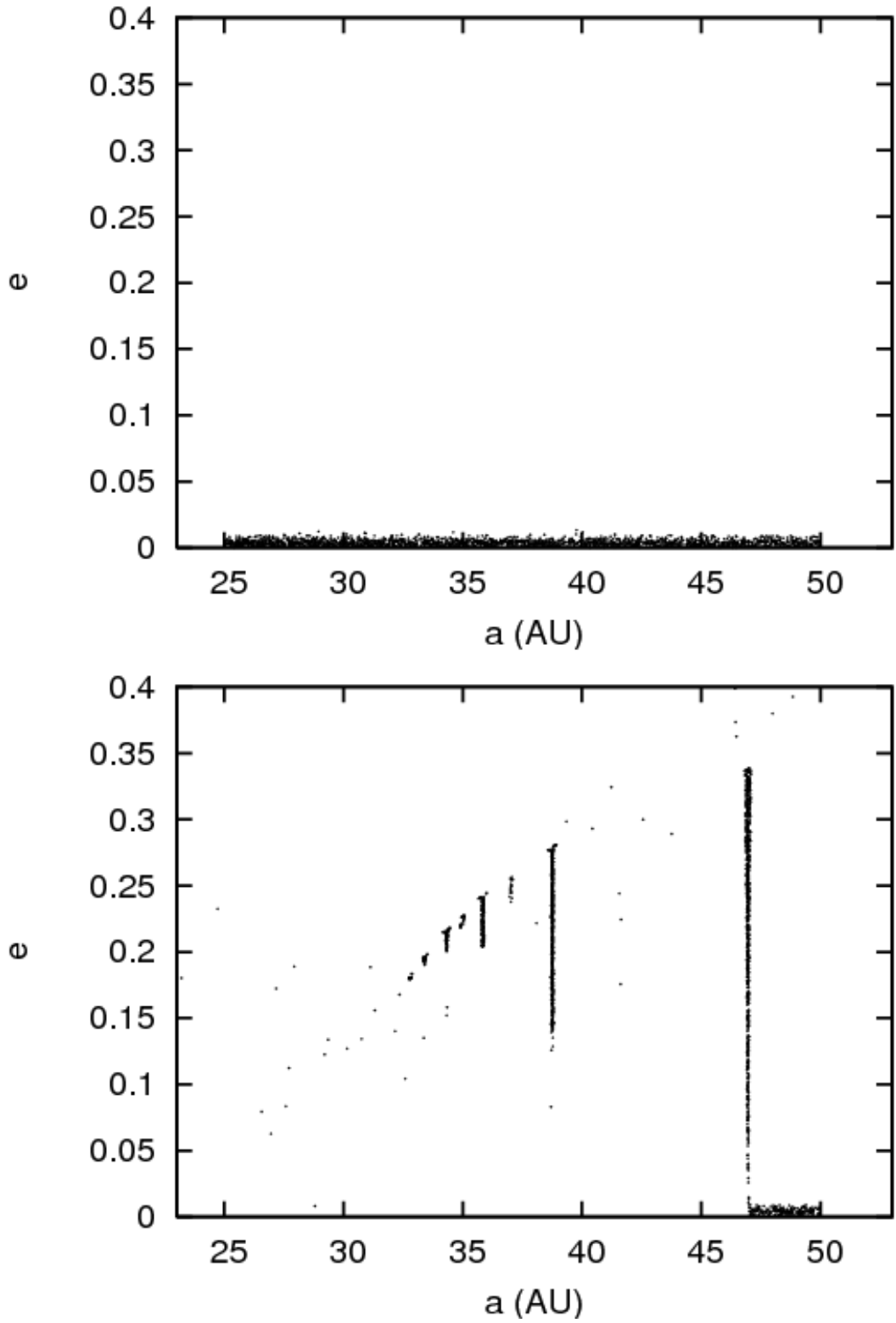}

\caption{The a-e plot of the initial planetesimals disk (top) and the disk after sweeping by MMR of Neptune 
(bottom). The resonant 
capture process is effective and the 100 \% of the particles which encounter the 2:3 or 1:2 MMR are 
captured into the resonances.}
\label{snapshot_initial}
\end{figure*}

The figure \ref {snapshot_initial} shows our initial disk and the result after Neptune migration.
Using cold disk, the resonant capture is very effective and all particles swept by the 1:2 and 2:3 MMR of
Neptune are captured into the resonance.
The perturbation from
other planets reduce the capture probability and this is some ideal results. But this condition is good for
investigate the genuine effect of the self gravity. 
From now on, we focus on the planetesimals which captured into the 1:2 MMR.

We pick up 1000 (model 1, 6, 7, 8), 2000 (model 2,4), 500 (model 3, 5) particles from the 1:2 MMR population and
assign mass.
We define the planetesimals which captured in the 1:2 MMR as those whose semi-major axis is $46.7~AU<a<47.7~AU$.

In the model 1, The total mass of Twotinos is about 0.1 $M_{\oplus}$ this correspond to the surface mass density 
which is about 100 times smaller than that of the minimum-mass solar nebular model (Hayashi et al 1981).
The diameter of each particle is about 450 km.
The detailed parameters are shown in Table \ref {initial_conditions}. In model 2-5, we change the total mass and
the mass of planetesimals to see the dependence of the relaxation time on these 
parameters. In model 6, 7, we change the inclination 
dispersion to see the effect of inclination. We make these models by changing the dispersion of inclination when
the Neptune migration is over.
The model 8 is the model of massless case
and it is used for comparison with other models.  
We integrate the system for $1.6\times 10^7$ years considering the gravitational interaction
between the planetesimals.
The gravity to Neptune from planetesimals are neglected to see the genuine internal gravitational effect.

\begin{figure*}
\includegraphics[width=100mm]{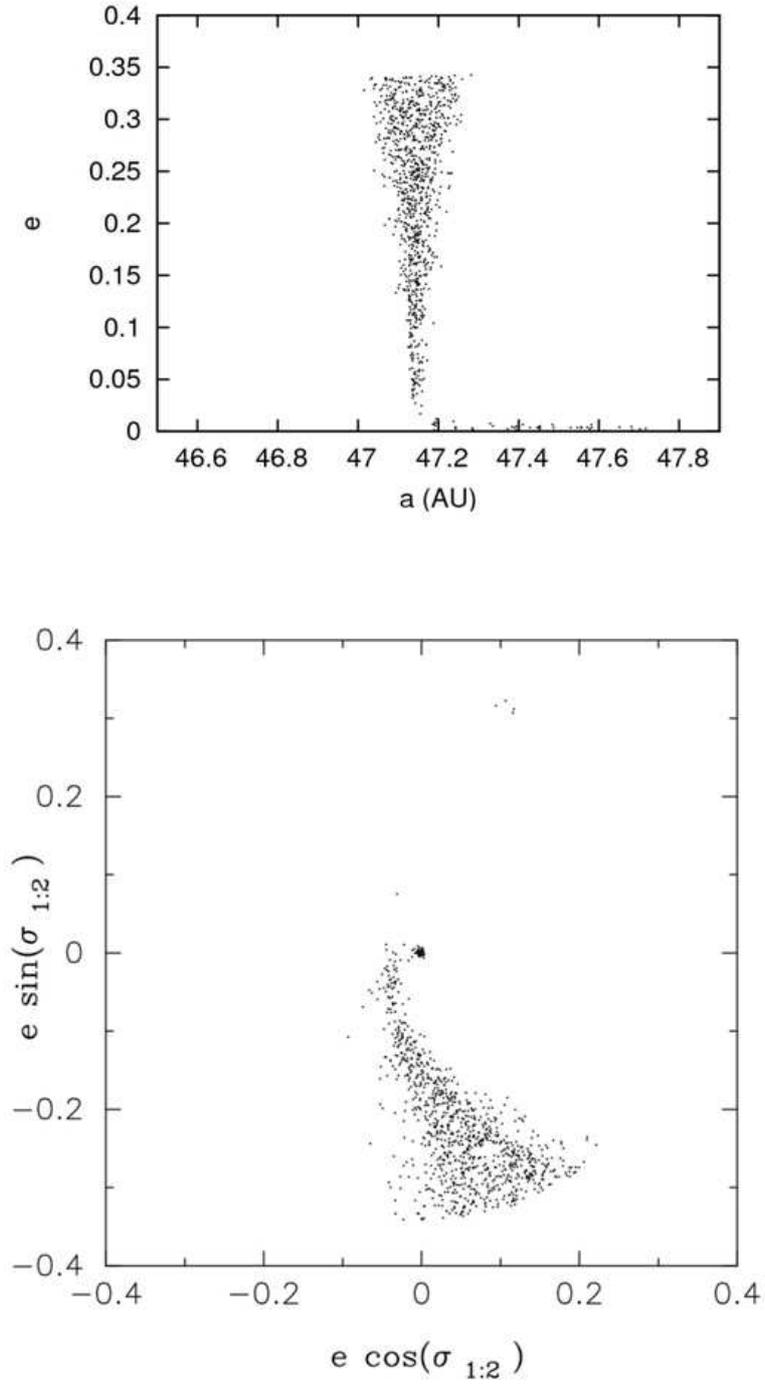}

\caption{The a-e plot (top) and $e\cos(\sigma_{1:2})-e\sin(\sigma_{1:2})$ plot (bottom) of the initial condition
for our simulations, where $\sigma_{1:2}$ is the resonant angle $\sigma_{1:2}=2\lambda-\lambda_N-\varpi$.}
\label{snapshot_initial_resonance}
\end{figure*}

The figure \ref {snapshot_initial_resonance} shows the initial distribution of planetesimals
on $a$-$e$ plane and $e\cos(\sigma_{1:2})$-$e\sin(\sigma_{1:2})$ plane, where $\sigma_{1:2}$ is
the resonant angle, $\sigma_{1:2}=2\lambda-\lambda_N-\varpi$. 
Most particles are trapped around $\langle \sigma_{1:2}\rangle \simeq 270^\circ$.
 This asymmetric capture is agree well with the result of Chiang \& Jordan (2002).

\begin{table*}[hbt]

\caption{The initial conditions}		
\label{initial_conditions}
\begin{center}
\begin{tabular}{cccccc}
\hline\hline
 Model & {\it N} & {\it M} (g) &{\it r} (km) & total mass ($M_{\oplus}$)  & $\langle i^2\rangle^{1/2}$\\
\hline
 1 & 1000 & $ 7.62 \times 10^{23} $ & 450 & 0.127 & 0.002 \\
 2 & 2000 & $ 3.81 \times 10^{23} $ & 360 & 0.127 & 0.002\\
 3 & 500 & $ 1.52 \times 10^{24} $ & 570 & 0.127  & 0.002\\
 4 & 2000 & $ 7.62 \times 10^{23} $ & 450 & 0.254 & 0.002 \\
 5 & 500 & $ 7.62 \times 10^{23} $ & 450 & 0.0635 & 0.002\\
 6 & 1000 & $ 7.62 \times 10^{23} $ & 450 & 0.127 & 0.01  \\
 7 & 1000 & $ 7.62 \times 10^{23} $ & 450 & 0.127 & 0.02 \\
 8 & 1000 & 0 & 0 & 0 & 0.002  \\
\hline

\end{tabular}
\end{center}
\end{table*}

\section{Results}
\label{results}
\begin{figure*}
\includegraphics[width=150mm]{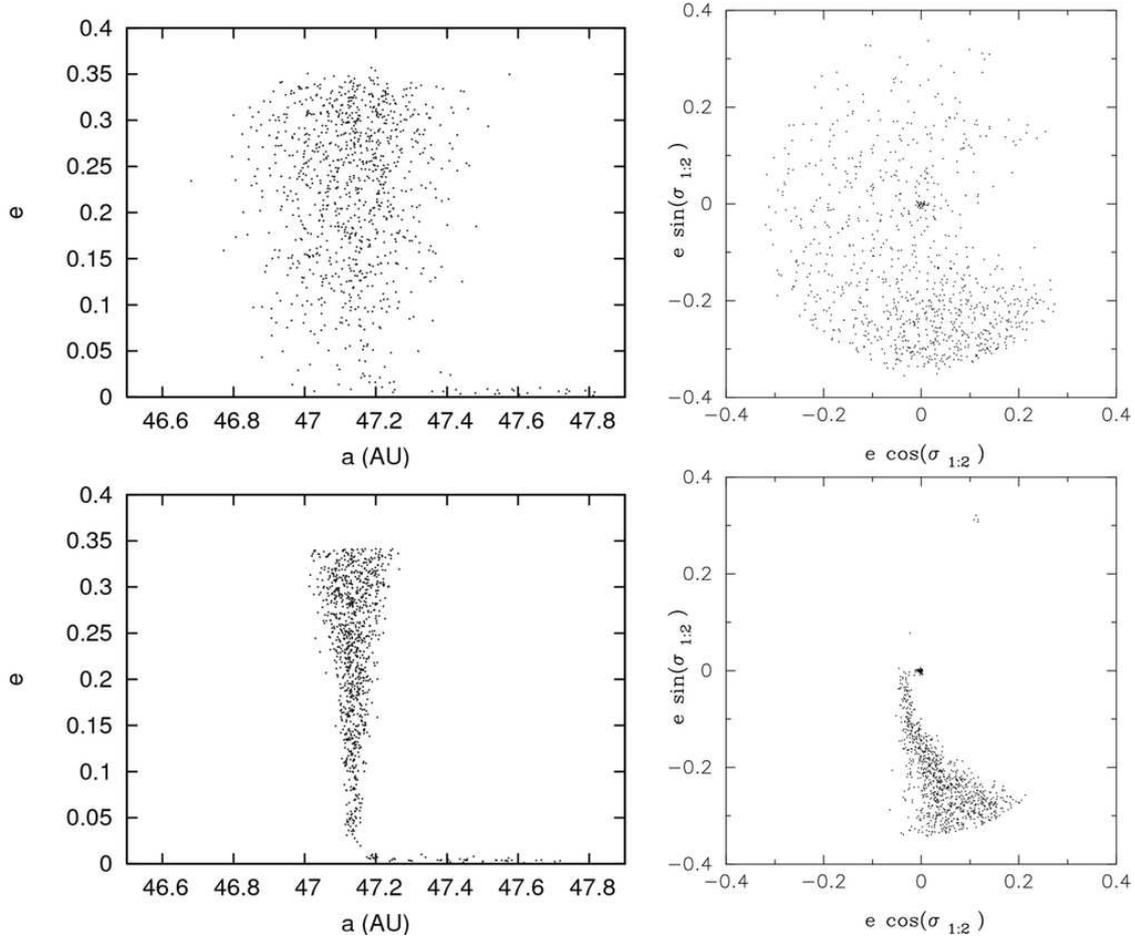}

\caption{The planetesimals distribution at $t = 1.6\times
  10^{7}$ years. The top left and the top right are plots on a-e plane and 
$e\cos(\sigma_{1:2})$-$e\sin(\sigma_{1:2})$ plane of model 1, respectively.
The bottom left and bottom right are the same plots of model 8 (the massless case).}
\label{snapshot_after_integral}
\end{figure*}

Figure \ref{snapshot_after_integral} shows the snapshot of model 1 and model 8 (massless case) at $t=1.6 \times 10^{7} $ years.
The top right panel shows the snapshot on $e\cos(\sigma_{1:2})$-$e\sin(\sigma_{1:2})$ plane 
of the model 1. The particles distribute broad region. 
This means the libration of resonant angle are amplified by the self gravity. 
Because the amplitude of resonant angle correlate with the semi-major axis, the amplitude of semi-major axis
is also amplified as shown in the top left panel of the figure. Compared with the figure 5 of Nesvorn\'y and Roig (2001) 
and the top left panel of the figure, We can see that many planetesimals become unstable.
On the other hand, we can not find any obvious change in the result of model 8 (bottom panels) over the integration time.

\begin{figure*}
\includegraphics[width=150mm]{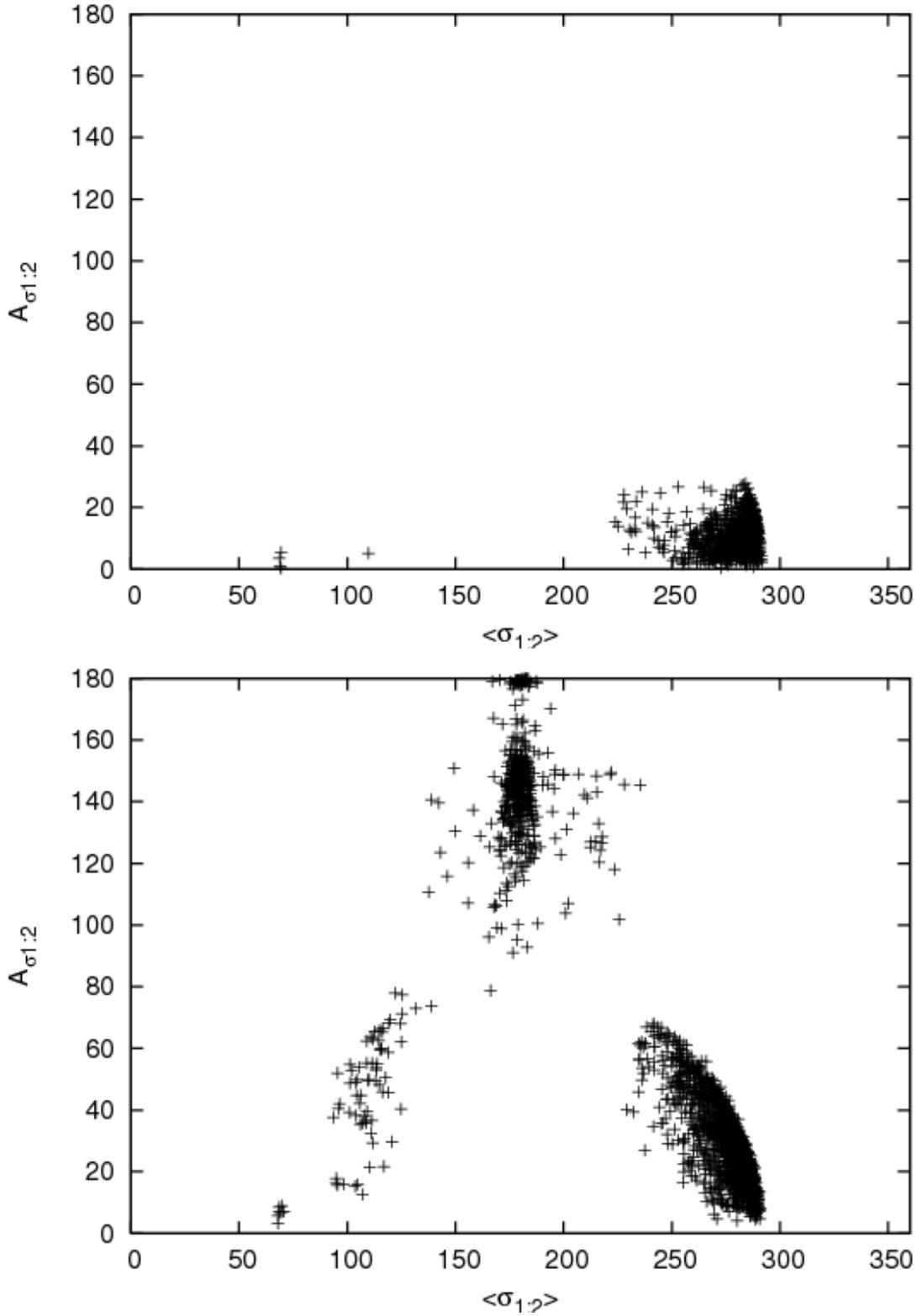}

\caption{The planetesimals distribution on $\langle \sigma_{1:2} \rangle$-$A_{\sigma_{1:2}}$ plane
at $t = 1.6\times10^{7}$ years is shown.
The top and bottom panel shows Model 8 and 1 respectively.
the mean value, $\langle \sigma_{1:2} \rangle$ is averaged over 160,000 years and
the amplitude is defined as 
$(\sigma_{max}-\sigma_{min})/2$, where $\sigma_{max}$ and $\sigma_{min}$ are
 maximum and minimum value of resonant angle in each 160,000 year.}
\label{snapshot_res}
\end{figure*}

We show the plot of the planetesimals on $\langle \sigma_{1:2} \rangle $ - $A_{\sigma_{1:2}}$ plane 
of the resonant angle in figure \ref{snapshot_res}.
We take snapshot each 1,600 year and the mean value, $\langle \sigma_{1:2} \rangle$ is 
averaged value of the resonant angle over 160,000 year. The amplitude, $A_{\sigma_{1:2}}$ is 
defined as $(\sigma_{max}-\sigma_{min})/2$. The $\sigma_{max}$ and $\sigma_{min}$ are
maximum and minimum value of resonant angle each  160,000 year. 
The "window time", 160,000 year is 1-10 times longer than the
period of oscillation of $\sigma_{1:2}$ and it is enough to estimate the mean, maximum, minimum values.
The bottom panel shows the result of model 1 and we can see the three populations. The two populations whose amplitudes are
less than 80 deg are correspond to the populations whose orbits are
asymmetric and tadpole like on $e\cos(\sigma_{1:2})$-$e\sin(\sigma_{1:2})$ plane and the population  whose 
$\langle \sigma_{1:2} \rangle$ is around $180^\circ$ is the population whose orbits 
 is symmetric and horse-shoe like orbit.
The existence of these three population is consistent with the result of Chian and Jordan (2002) (see, figure 7 of their paper).
The result of massless case (top panel) does not show any obvious change.

Note that The planetesimals distribution which have $\langle \sigma_{1:2} \rangle\simeq 180^\circ$ and $A_{\sigma_{1:2}} \simeq 180^\circ$
 on top panel is those which is ejected from 1:2 MMR.
The planetesimals distribution which have $220^\circ< \langle \sigma_{1:2} \rangle<250^\circ$ and $A_{\sigma_{1:2}} \simeq 20^\circ$
on bottom panel is those which have low eccentricity ($e<0.1$).
For those planetesimals, the two islans on $e\cos(\sigma_{1:2})$-$e\sin(\sigma_{1:2})$ plane
 are around $\sigma_{1:2} \simeq 250^\circ,~70^\circ $ (see, top right panel of figure 6 of Murray-Clay \& Chiang (2005) ).

\begin{figure*}
\includegraphics[width=150mm]{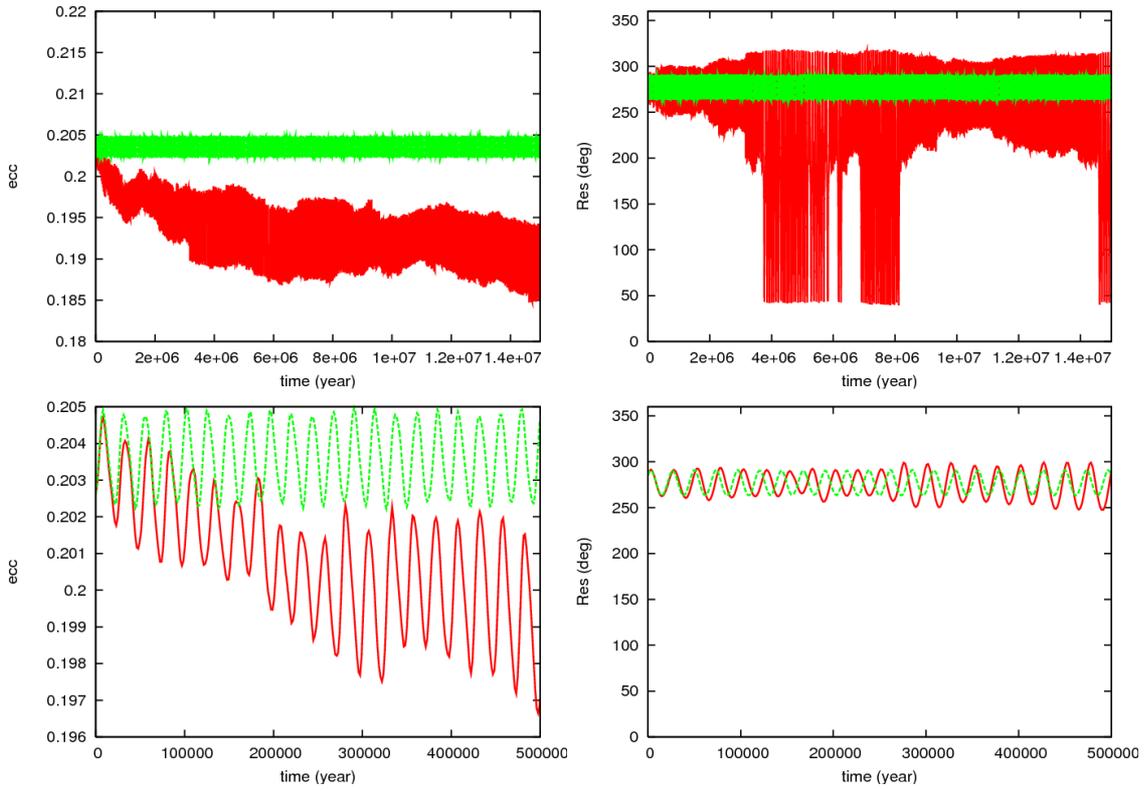}

\caption{The time evolution of the resonant angle and eccentricity of a planetesimal.
The top panels shows the long term evolution of the eccentricity (left) and the resonant 
angle (right). The time range of top panels is $[0:1.5 \times 10^{7}]$ years.
The bottom panels shows the short range evolution of the the eccentricity (left) 
and the resonant angle (right). The time range of the bottom panels is $[0:5 \times 10^{5}]$ years}
\label{time_orbital}
\end{figure*}

We show an example of the time evolution of the eccentricity and the resonant angle of a planetesimal 
with (red line, model 1) and without (green line, model 8) self gravity in figure \ref{time_orbital}. 
The random walk nature is seen in the change of the eccentricity of red line. 
The top right panel shows the time evolution of the 
resonant angle. Around $4\times 10^6$ year, the amplitude of resonant angle 
changes discontinuously for the red line. This shows the orbit of the particle on the 
$e\cos(\sigma_{1:2})$-$e\sin(\sigma_{1:2})$ plane transit from tadpole orbit to horse-shoe orbit. 
The transit from tadpole orbit to horse-shoe orbit
and vice-vase occur several time. This shows that the particles can easily across the
separatrix due to self gravity. This nature is also displayed in the Chiang and Jordan (2002). Thus, this is
also arisen by the secular perturbation from Planets.
 
The important difference between the effect of the self gravity
and that of the secular perturbation from the Planets is that the self gravity can amplify the libration of resonant angle 
even though the particles initially have very small amplitude such as $A_{\sigma_{1:2}} <30^\circ$. 
As Nesvorn\'y and Roig (2001) have shown,
the change of the amplitude of resonant angle which is initially $A_{\sigma_{1:2}} <30^\circ$ 
practically unchanged by the secular perturbation of the planets for the age of Solar System.  
On the other hand, the amplitude increases by the self gravity even though particles
initially have low amplitude.
Figure \ref{snapshot_initial_resonance} shows all particles initially satisfy the condition, 
$A_{\sigma_{1:2}} <30^\circ$.

\begin{figure*}
\includegraphics[width=150mm]{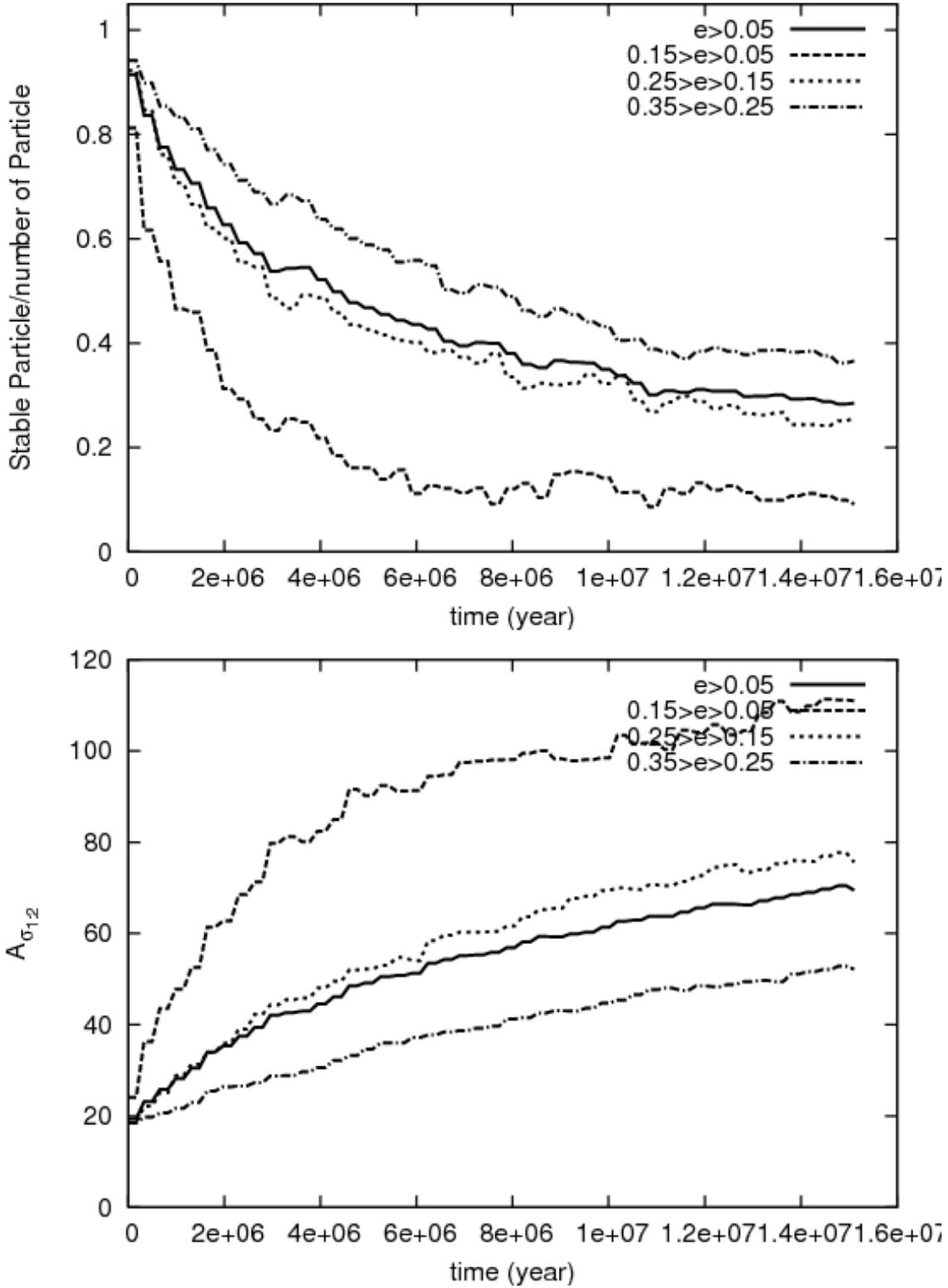}

\caption{The time evolution of the fraction of the "stable" particles which does not satisfy
the unstable condition (\ref{unstable_condition})(top) and the averaged value of 
the amplitude of resonant angle (bottom). The average is taken with respect to the 
particles which have similar eccentricity.  The solid, dashed, dotted, dot-dash line show
the results of $0.05<e$, $0.05<e<0.15$, $0.15<e<0.25$ and $0.25<e<0.35$, respectively.}
\label{time_fraction}
\end{figure*}

The time evolution of the fraction of the "stable" planetesimals in Model 1
is shown in the top of figure \ref{time_fraction}. 
We refer the condition, 
\begin{eqnarray}
\label{unstable_condition}
A_{\sigma_{1:2}} > 30^\circ
\end{eqnarray}
as "unstable condition" and "stable" planetesimals means 
the planetesimals which does not satisfy this condition.
As Nesvorn\'y and Roig (2001) insists,
the chaotic evolution of the planetesimals which have the large amplitude of resonant angle 
becomes very strong and many of them are expected to escape from the 1:2 MMR. The condition, 
$A_{\sigma_{1:2}} > 30^\circ$ can be used as the crude threshold to determine whether planetesimal 
is unstable or not.  
The fraction of stable particles is calculated with respect to 
the particles which have $0.05<e$, $0.05<e<0.15$, $0.15<e<0.25$, 
and $0.25<e<0.35$. At first, each region roughly contains 920, 140, 320, 460, 
particles, respectively 
and these numbers hardly change during the integration.
For the planetesimals in the low eccentricity region ($0.05<e<0.15$), The fraction of the stable particles 
quickly decreases. As the eccentricity increases the change of the fraction becomes 
slower. This comes from the difference of velocity dispersions. 
The bottom of the figure \ref{time_fraction} shows the time evolution of the 
averaged amplitude of resonant angle.
The average is taken over the particles at each eccentricity region. 
The resonant angle of particles in the
low eccentricity region increases quicker than that of high eccentricity.

The solid line of the top panel shows that more than half of 
planetesimals in the 1:2 MMR satisfy the unstable 
condition within the $4.0 \times 10^6$ years in model 1.
This value can not be directly applied to real 
primordial TNOs because the parameter of our simulation is different from the 
real TNOs. Thus, how this result can extrapolate to realistic case must be discussed.

\begin{figure*}
\includegraphics[width=150mm]{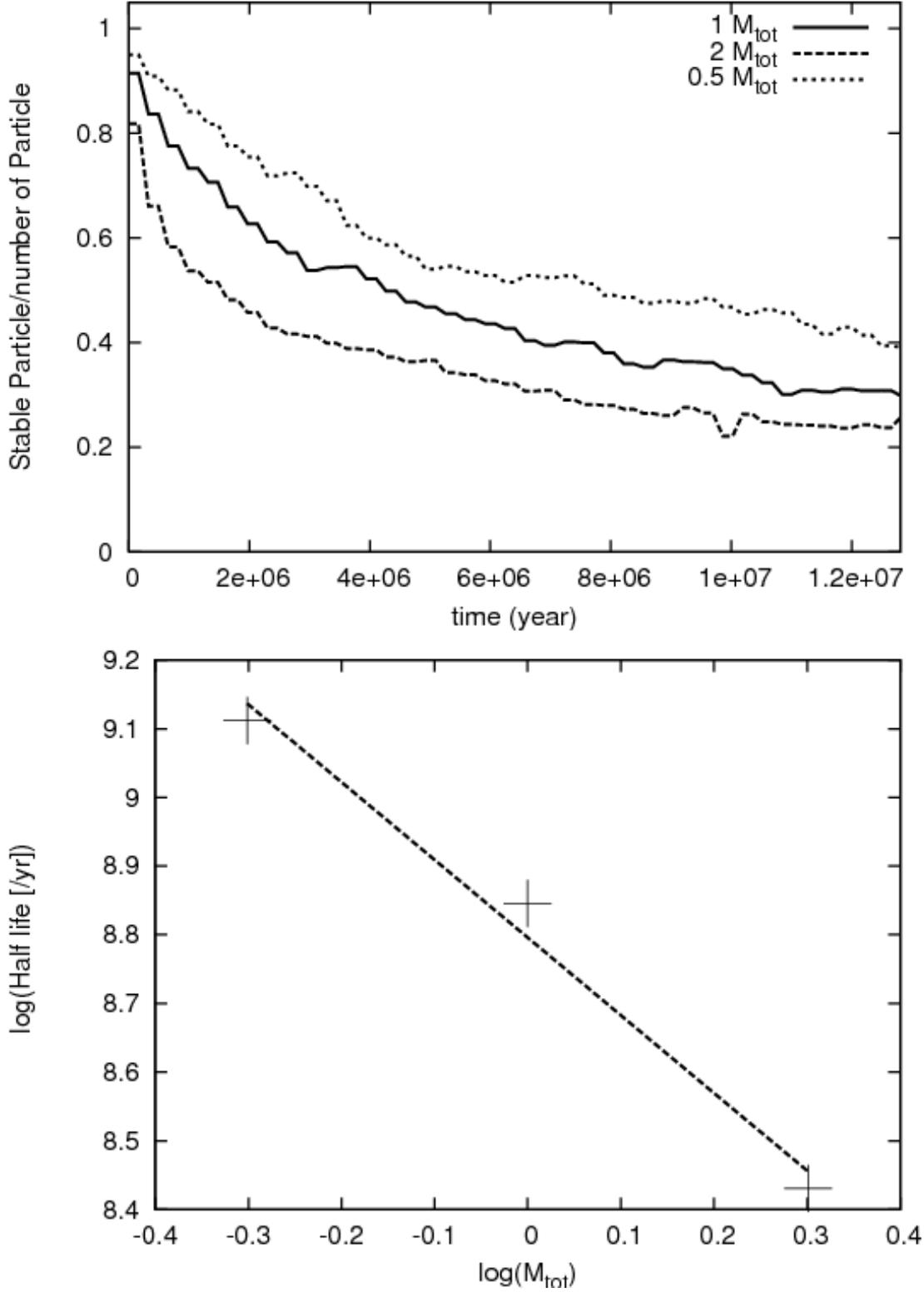}

\caption{
The top panel is the same as the top of figure \ref{time_fraction} but the solid, dashed, dotted line are 
the result of  model 1, model 4, model 5.
The total mass of each model is $1 M_{tot}$ (model 1, solid), $2 M_{tot} $ (model 4, dashed), $0.5 M_{tot}$ 
(model 5, dotted), respectively and other parameters are the same. 
the results obtained using the particles whose eccentricity is $e>0.05$.
$M_{tot}$ is the total mass of model 1 and its value is about $0.1 M_\oplus$. The
bottom panel shows the result of linear regression. We plot the total mass of Twotinos vs the "half-life period" 
in log scale. The gradient of the linear curve $a$ is $a = -1.13152 \pm 0.1426~(12.6\%)$}
\label{timefraction-surface}
\end{figure*}

The top of the figure \ref {timefraction-surface} shows
how the fraction depends on the total mass of Twotinos. We vary the total mass as $2M_{tot}$ (model 2),
$M_{tot}$ (model 1),$0.5 M_{tot}$ (model 3) where $M_{tot}$ is the total mass of the
model 1 and its value is about $0.1~ M_{\oplus}$. 
The particles which have $e>0.05$ are used to calculate the fraction. In the bottom of the figure 
\ref{timefraction-surface}, we plot the total mass vs the "half-life period" 
which is the time in which more than 50 \% of particles satisfy the unstable condition in logarithm scale. 
We fit the points with the linear curve $\log_{10} y=a \log_{10} x + b$ 
by least-squares method and obtain the value of  the gradient, $a = -1.13152 \pm 0.1426~(12.6\%)$. 

\begin{figure*}
\includegraphics[width=150mm]{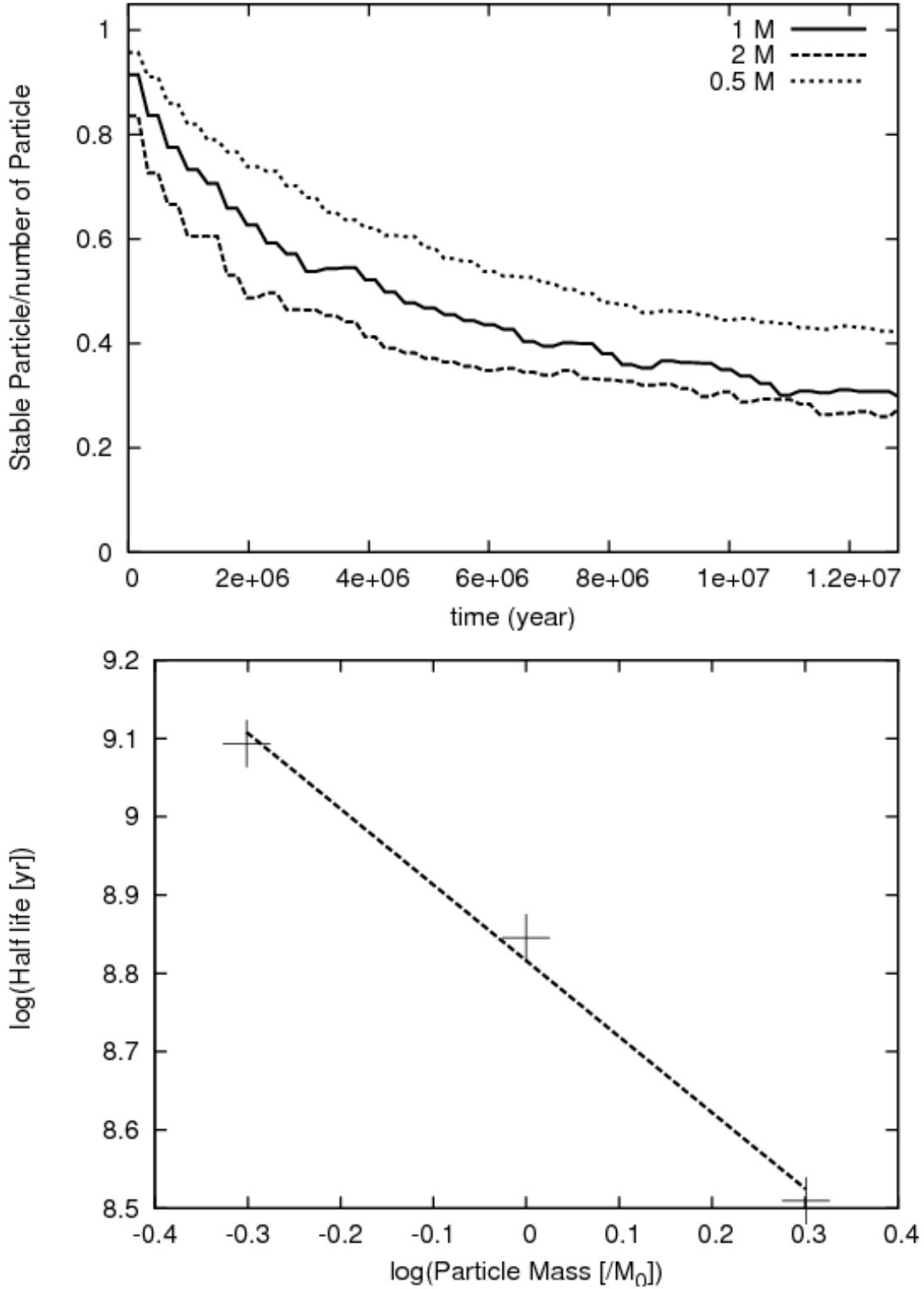}

\caption{The top panel is the same as the top of figure \ref{time_fraction} but the solid, dashed, dotted line are 
the results of model 1, 2, 3, respectively. The mass of planetesimals of each model is $M_0,~2M_0,~0.5M_0$, respectively
and other parameters are the same. The results obtained using the particles whose eccentricity is $e>0.05$.
The $M_0$ is the mass of particle of model 1 and its value is about $7.62\times 10^{23}$ g.
The bottom panel shows the result of linear regression. We plot the mass of each planetesimal
 vs the "half-life period" in log scale. The gradient of the linear curve $a$ is $a = -0.9693 \pm 0.0844 (8.707\%)$. 
}
\label{timefraction-mass}
\end{figure*}

The figure \ref{timefraction-mass} shows the dependency on the mass of the particles of the half-life period. 
The values of the mass is $2.0M$ (model 5), $1.0M$ (model 1), $0.5M$ (model 4), 
respectively. Where $M_0$ indicate the
mass of each planetesimal of model 1 and its value is about $ 7.6\times 10^{20} $ kg.
The bottom of the figure \ref{timefraction-mass} shows the $ \log ({\rm Mass}) $-
$\log ({\rm halflife})$ relation. We fit the points same as above and the gradient is 
$a = -0.9693 \pm 0.0844 (8.707\%)$. These results are good agreement with the general feature of the relaxation
timescale for the gravitational interaction, $t_{relax} \propto (m_pM_{tot})^{-1} $.
Thus, we conclude the relaxation time of the amplitude of resonant angle also proportional to $(m_pM_{tot})^{-1}$. 

\begin{figure*}
\includegraphics[width=150mm]{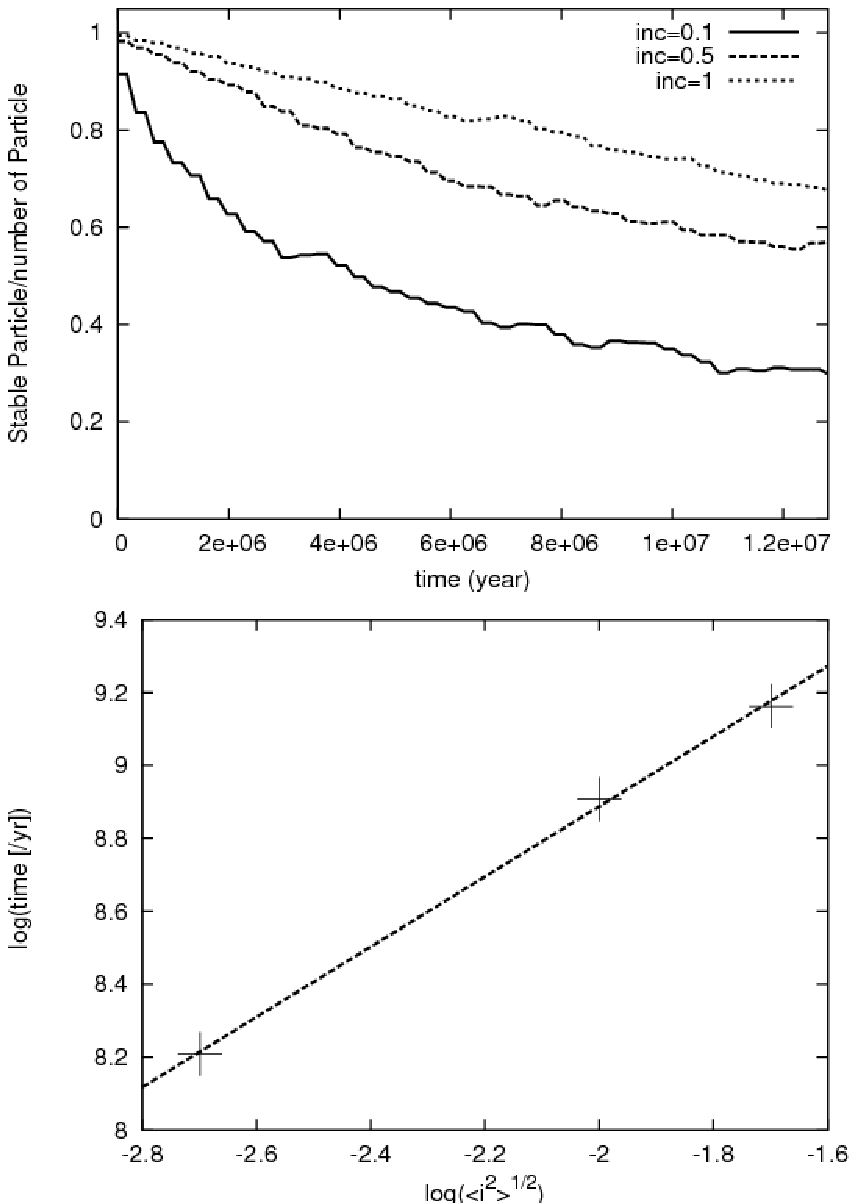}

\caption{The top panel is the same as the top of figure \ref{time_fraction} but the solid, dashed, dotted line are 
the results of model 1, 6, 7. The inclination dispersion, $\langle i^2\rangle^{1/2}$ of each model
is 0.002, 0.01, 0.02, respectively and other parameters are the same.
The results obtained using the particles whose eccentricity is $e>0.05$.
The bottom panel shows the result of linear regression. We plot the inclination dispersion vs the time in which the 25\%
of particles become unstable in log scale. The gradient of the linear curve $a$ is $a = 0.962302 \pm 0.03508~(3.645\%)$.
}
\label{time_inc}
\end{figure*}

The figure \ref{time_inc} shows how the timescale is changed by the inclination dispersion. 
The planetesimals whose eccentricity is higher than 0.05 are considered.
The each line in the top panel shows the cases the dispersion, $\langle i^2 \rangle^{1/2}$ are
0.002, 0.01 and 0.02, respectively.  
We plot the inclination dispersion vs the time at which 25 \% of the particles satisfy the unstable conditions 
in logarithm scale in the bottom panel. the gradient, $a = 0.962302 \pm 0.03508~(3.645\%)$. 
With the above results, we conclude that the half-life period is proportional to $(M_{tot} M)^{-1}\langle i^2\rangle^{1/2}$.
Note that the dependency for inclination is not theoretically supported. Thus, the dependency for inclination
is just phenomenalistic result and is not justified for all parameter space. But our extrapolation for inclination is up to
one order of magnitude and our claims will be justified.

Our results can be summarized in the formula,
\begin{eqnarray}
\label{estimate_fourmular}
t_{half}=4\times10^6(\frac{M_{tot}}{0.1M_\oplus})^{-1}(\frac{M_p}{7.6\times10^{23} g})^{-1}
(\frac{\langle i^2\rangle^{1/2}}{0.002}) (years).
\end{eqnarray}
We can estimate the unstable time-scale for realistic systems using this relation and 
the results we show above. 

At first, we apply our result to the parameter of current Twotinos.
Trujillo {\it et al.} (2001) estimate that the total mass of CTNOs is about 0.03 $M_\odot$ and
the population ratio of the classical TNOs and Twotinos are 1.0:0.07. 
They also pointed out that the differential size distribution of CTNOs is power law with 
exponent, $ q \simeq 4.0$. This correspond to the power law differential mass distribution whose exponent is about 2.0.
In this case, the main contribution of gravitational effect is come from the maximum size.
If we assume the mass distribution is same between CTNOs and Twotinos,
the total mass of Twotinos is about 0.002 $M_\oplus$ and $m_p$ can be assumed to be maximum mass of Twotinos.

The maximum size of Twotinos is not well understood.
We assume the maximum diameter of Twotinos is about 400 km, 
this is equal to the estimated diameter of largest Twotinos, ${\rm 2002 WC_{19}}$. 
The inclination dispersion of Twotinos assumed to be 5 deg.
From eq \ref {estimate_fourmular}, we can estimate the half-life period of real Twotinos 
is about $3\times10^{11}$ years. This value is very large compared to the age of Solar system.
Thus, we conclude that the self gravity effect is not important at the present day.

Next, we consider more drastic case in which the total mass of primordial Twotinos is about $3 M_\oplus$. 
This is the similar case Levison and Morbidelli (2003) assumed and the natural outcome if the planetesimals
initially distribute up to 47 AU with the power law density profile of minimum-mass solar nebular model (Hayashi 1981) and 
Neptune migrated very smoothly.
We assume that the largest size of Twotinos is still about 400 km. This is very modest assumption. The inclination
is assumed to be order 1 deg again. This is also very modest assumption for bulk of primordial Twotinos because it
is very rare to gain high inclination during migration (Gomes 2000). 
The power law exponent is assumed to be $q\simeq4.0$. 
Then, the half-life period of this case become about $10^6$ year. This is
sufficiently small compared to the age of Solar system. We can crudely estimate the rate of the particles which
remain the resonance for the age of Solar system using this result. The half of particles in the resonance 
becomes unstable within the first $10^6$ year and the total mass of stable planetesimals roughly
halves and the half-life period doubles, if we neglect the contribution which come into stable region.
In the successive $2 \times 10^6$ years, the half of particles in the
stable region becomes unstable, thus totally 75 \% of particles becomes unstable for $3 \times 10^6$ years.
By iterating this process the mass of Twotinos can be reduce until the half-life 
period become comparable to $10^9$ years.
This is significant result and invoke the reexamination to previous works which try to explain 
the dynamical structure of TNOs. Note that this kind of orbital instability never occur by the secular
perturbation alone because the self gravity can amplify the libration of the resonant angle
of the particles which initially have small amplitude (Nesvorn\'y \& Roig 2001).

From above estimate, We conclude that the total mass of Twotinos decreases until the order 
of $0.01~M_{\oplus}$ even if Twotinos initially contain huge mass because the effect of self gravity
is proportional to the total mass and the total mass decreases until the half-life period
becomes comparable to the age of solar system. This estimate has large uncertainty because of the
lack of sufficient survey of Twotinos.

If the index of the size distribution, q is proven to be larger than 4 for Twotinos, the largest 
objects do not contain the majority of the mass.
In this case, the total mass must be adjusted down to include only the mass in those maximum mass planetesimals 
and the half-life period increases. Thus, the estimated value of timescale can change depending
on the value of exponent, $q$ and the $M_{tot}$ should be interpreted as total mass of large bodies.
The value of $q$ is not well determined. For example, Petit {\it et al.} (2006) estimate the power 
law index for TNOs, $q\simeq 4.8$ and much larger than the result of Trujillo {\it et al.} (2002).
Recently, theoretical work about the size distribution of TNOs have done and it suggests that
$q\simeq 4.0$ (Sclichting \& Sari 2010). Further observational and theoretical work is required to
determine the half-life period more exactly. 

We point out the possibility that the high existing rate of high inclination population in 1:2 MMR is 
stem from the selective escape of the low inclination bodies. During orbital migration of Neptune, some of primordial Twotinos may
gain high inclination (say, about 10 deg) with the similar effect proposed by Gomes (2000). This mechanism is very rare and it 
seems to be difficult to explain the large inclination of Twotinos with this mechanism alone.
But if there was huge amount planetesimals between 30 AU and 47 AU, the mechanism can provide sufficient number of planetesimals 
to account for the present high inclination population of Twotinos. 
After the migration, low inclination bodies selectively escape from 1:2 MMR because half-life period of low inclination body 
is an order of magnitude lower than high inclination body. Consequently, high inclination bodies selectively remain at 1:2 MMR. 
This can be a explanation why most of planetesimals of 1:2 MMR have high inclination. 

\section{Summary and Discussion}
\label{discussion}
In this paper, we showed the results of the simulations which investigated the self gravity effect on Twotinos.
We confirm the self gravity does change the resonant angle of planetesimals. 
We investigated how the timescale in which Twotinos becomes unstable depend on the total mass of Twotinos, 
the mass of planetesimals, and the inclination dispersion. The timescale obey the formula,
\begin{eqnarray}
t_{half}=4\times10^6(\frac{M_{tot}}{0.1M_\oplus})^{-1}(\frac{m_p}{7.6\times10^{23} g})^{-1}
(\frac{\langle i^2\rangle^{1/2}}{0.002}) (years),
\end{eqnarray}
Here, $M_{tot}$ is interpreted as the total mass at large sizes and $m_p$ is the mass
of large bodies. $\langle i^2 \rangle^{1/2}$ are the inclination dispersion.

 Using this formula, we concluded that the total mass of Twotinos 
can be reduced to the order of $0.01~M_{\oplus}$ even if there was huge mass such as several order of earth 
mass in the 1:2 MMR because the effect of self gravity is proportional to the total mass
and sufficiently strong until the total mass fall bellow the threshold mass at which the timescale becomes
 comparable to the age of solar system.
These results invoke reexamination to many previous works such as the dynamical evolution model 
of TNOs (Levison and Morbidelli 2003) or Neptune migration speed estimated from the current 
orbital population of Twotinos (Murray-Clay \& Chiang 2005). 

The work of Levison \& Morbidelli (2003) relies on the mass carrying effect of the 1:2 MMR to explain 
the structure of the TNOs region. 
With the results done by Gomes (2003), their result successfully explain the important properties 
of TNOs. 
The model assumes that the total mass of 1:2 MMR population is about $3 M_\oplus$ when Neptune is at 22 AU and secular
resonance between Twotinos and Neptune induce the oscillation of eccentricity of Twotinos. When the planetesimals
reach small eccentricity, Neptune's stochastic migration kick out these bodies from resonance. This is the 
origin of the cold population in CTNOs. They also argue that almost 1:2 MMR population also kicked out during
the migration process by the stochastic migration.
But this argument contains self-contradiction. As many works have shown 
(Hahn \& Malhotra 1999, Zhou {\it et al}. 2002),
if Neptune's migration was much jumpy to shake off the bulk of the planetesimals in the 1:2 MMR, 
it seems to impossible to capture the planetesimals into the 1:2 MMR. Further more, Murray-Clay \& Chiang (2006)
pointed out that primordial planetesimal disk of TNOs region did not have big objects which can induce jumpy migration. 
This seems to be a serious problem of this model.

On the other hand, our results naturally explain the removing process from 1:2 MMR after Neptune
migration finished. Further more the particles which has low eccentricity becomes unstable quicker 
than high eccentricity particles as we have shown. It can be an alternative mechanism to kick 
out the planetesimals into classical TNOs region.
Thus it is unnecessary to assume the strongly junky migration if the self gravity effectively work 
and the model suggested by Levison \& Morbidelli (2003) becomes more attractive.
Of course further research is needed to determine how the self gravity acts during the Neptune migration.

If the huge mass such as several order of earth mass is really captured and released from Twotinos, 
these particles may affects the additional migration to Neptune or other planets as Morbidelli {\it et al.} 
(2007) pointed out. This effects can change the entire structure of trans neptunian region and the scenario
of the formation of our Solar system.
To investigate the "second migration process" effects and how it will change the entire structure of 
trans Neptune region is beyond our scope in this paper and further investigations are required.

Murray-Clay \& Chiang (2005) estimate the migration speed of Neptune from the current population of 1:2 MMR.
Their result show whether the planetesimals distribute on the "trailing island" or "leading island" whose liberation
center of oscillation of resonant angle is greater than $\pi$ as opposed to less than $\pi$  sensitively depends on
the timescale of Neptune migration. However, our results show that the planetesimals can "penetrate" from these island
due to the self gravity and destroy the information of the primordial dynamical structure of Twotinos.
Thus, we think more precise treatment is required to estimate the timescale of Neptune migrations from 
the population of Twotinos.

Our result is very strong because there is no arbitrary assumption and is natural outcome of physical principle.
More investigations are required whether our results can be the key mechanism to explain the entire
 structure of trans neptunian region but we believe our result play very important role for the orbital evolution
of TNOs.

\section *{Acknowledgments}
We thank Masaki Iwasawa, Eichiro Kokubo and Junichiro Makino for fruitful discussions 
This research is partially supported 
by the Special Coordination Fund for Promoting Science and Technology (GRAPE-DR project).
Ministry of Education, Culture, Sports, Science and Technology, Japan.

\end{document}